\documentclass[conference]{IEEEtran}
\usepackage[top=0.7in,bottom=1.049in,left=0.68in,right=0.68in]{geometry}
\usepackage{cite}
\usepackage{amsmath,amssymb,amsfonts}
\usepackage{algorithmic}
\usepackage{graphicx}
\usepackage{textcomp}
\usepackage{xcolor}
\usepackage[subrefformat=parens]{subcaption}
\usepackage{multirow,bm}
\usepackage{booktabs}
\usepackage{colortbl}

\def\BibTeX{{\rm B\kern-.05em{\sc i\kern-.025em b}\kern-.08em
    T\kern-.1667em\lower.7ex\hbox{E}\kern-.125emX}}
\begin{document}

\title{Online Trainable Wireless Link Quality Prediction System using Camera Imagery}

\author{
\IEEEauthorblockN{
\normalsize Sohei Itahara\IEEEauthorrefmark{1},
\normalsize Takayuki Nishio\IEEEauthorrefmark{1},
\normalsize Masahiro Morikura\IEEEauthorrefmark{1}, and
\normalsize Koji Yamamoto\IEEEauthorrefmark{1}
}
\IEEEauthorblockA{
\IEEEauthorrefmark{1}\small Graduate School of Informatics, Kyoto University,
Yoshida-honmachi, Sakyo-ku, Kyoto 606-8501, Japan\\
Email: nishio@i.kyoto-u.ac.jp
}
}

\maketitle

\begin{abstract}
    Machine-learning-based prediction of future wireless link quality is an emerging technique 
    that can potentially improve the reliability of wireless communications, especially at higher frequencies (e.g., millimeter-wave and terahertz technologies), through predictive handover and beamforming to solve line-of-sight (LOS) blockage problem.
    In this study, a real-time online trainable wireless link quality prediction system was proposed; the system was implemented with commercially available laptops. 
    The proposed system collects datasets, updates a model, and infers the received power in real-time.
    The experimental evaluation was conducted using 5 GHz Wi-Fi, where received signal strength could be degraded by 10 dB when the LOS path was blocked by large obstacles.
    The experimental results demonstrate that the prediction model is updated in real-time, adapts to the change in environment, and predicts the time-varying Wi-Fi received power accurately.
\end{abstract}

\begin{IEEEkeywords}
mm-wave, machine learning, camera assisted, implementation, 5G, beyond 5G
\end{IEEEkeywords}

\section{Introduction}

Wireless communication technology has evolved over the years and is presently  one of the most advanced and sought-after technology.
However, the bandwidth of a wireless network is limited.
To overcome this well-established limitation,
 it is anticipated that the fifth-generation (5G) and beyond-5G wireless networks would
  attain higher-frequency bands, including 3--6 GHz, the millimeter wave (mmWave), and THz bands 
 (28 GHz, 60 GHz, and above 100 GHz)~\cite{Rappaport2019Wireless}.
These higher-frequency bands offer huge bandwidths and potentially enable data rates of several tens or hundreds of Gbit/s.
However, owing to the high attenuation and diffraction losses, 
the blockage of the line-of-site (LOS) path profoundly degrades the wireless link quality.
Specifically, in mmWave communications,
the received power suddenly drops by 20 dB or more when a LOS path is blocked by moving obstacles such as pedestrians or vehicles~\cite{collonge2004influence}.
Therefore, the link quality of the communication systems operating at higher frequency bands varies rapidly due to the LOS path blockage. 

To overcome this problem, vision-assisted radio frequency (RF) prediction~\cite{nishio2019proactive,ayvacsik2019veni}, learning-based proactive wireless control (handover~\cite{koda2019handover}, and beamforming~\cite{iimori2020stochastic}) approaches have been developed. 
These works leverage a camera to observe the communication environments and obtain information about moving obstacles that cause severe attenuation on the mmWave and THz link when they block the LOS path.
We have demonstrated that the time-varying mmWave received power can be predicted from camera imagery by applying a contemporary machine learning (ML) approach~\cite{nishio2019proactive}.
The vision-based RF prediction is a new paradigm that can potentially solve the blockage problems in mmWave and THz communications, and there are many challenges and opportunities.

The critical issue in developing a high-accuracy prediction model is that the model training process 
requires a large training dataset;
 moreover, the expected computation cost is high.
Existing studies~\cite{nishio2019proactive,koda2019handover} conducted offline model training. The training dataset was populated first; subsequently, an ML model was trained using the dataset.
Therefore, offline training causes a delay between the times when the data are obtained and when the model is updated, although the dataset of images and received power can be obtained and accumulated moment by moment.
The delay prevents the model from adapting to the real-time environment.
\begin{figure}[t]
  \centering
  \includegraphics[width=0.4\textwidth]{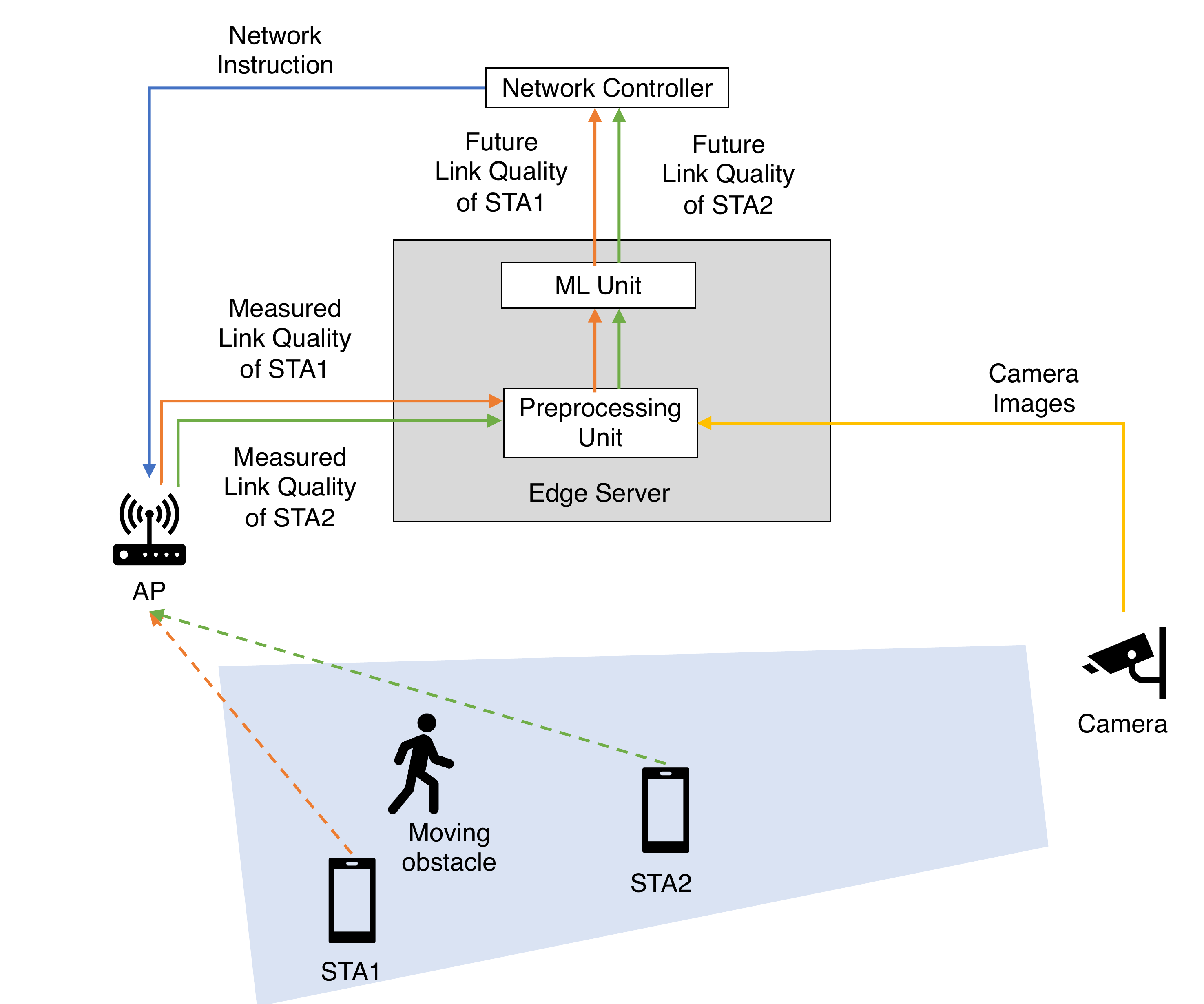}
  \caption{System model. Dotted lines indicate wireless links, whereas solid lines indicate wired links.}
  \label{fig:ideal_system}
\end{figure}

In this study, an online model training and prediction system was proposed for the image-based received power prediction.
The main contributions of this study are threefold: 
(1) design of an online training and prediction system of the image-based received power prediction that collects datasets, updates a model, and infers the received power in real-time;
(2) implementation of the online system with 5 GHz Wi-Fi received power prediction using a standard laptop equipped with a camera; and 
(3) demonstrating that the prediction model was updated in real-time, adapted to the change in environment, and accurately predicted the time-varying Wi-Fi received power.
Notably, the experiments using 5 GHz Wi-Fi instead of mmWave communications are still reasonable to evaluate the image-based received power prediction since the received signal strength can be degraded by 10 dB, similar to  mmWave communications, when the LOS path is blocked by large obstacles. 

The rest of this paper is organized as follows:
Section~\ref{sec:online} describes the proposed online prediction system.
Experimental setups and implementation are described in Section~\ref{sec:setup}.
Section~\ref{sec:results} provides experimental results, and Section~\ref{sec:conclusion} concludes this paper. 

\section{Online Training Framework for Image-based Received Power Prediction}
\label{sec:online}

\subsection{System Design}

Figure~\ref{fig:ideal_system} presents a system model based on the previous works~\cite{nishio2019proactive,koda2020communication}.
The system consists of an access point (AP), a camera, an edge server, a network controller, and $N$ stations (STAs).
The server is connected to the AP, camera, and network controller by a wired local area network.
The STAs are connected to the network via a wireless connection to the AP, and periodically receive and send packets.
The wireless link quality (e.g., received power, throughput, and packet loss rate) of an STA is degraded when obstacles, e.g., a moving human or car, block the LOS path between the AP and a particular STA.
This study considers the received power as the wireless link quality and amis to predict the received power.
The AP measures the received power of signals from a particular STA and sends the power information to the server.
The camera captures color images and sends them to the server.
We have assumed a simple case in which the LOS path between an STA and the AP always lies in the field of view of the camera. 
The server predicts the future received power for each STA using the camera images and current received power as inputs.
Using the predicted received power information,
 the network controller sends the appropriate command to the AP, such as a request for handover or a beamforming instruction.

\subsection{Online Training and Prediction Procedure}

The server contains an ML unit and a preprocessing unit, functioning parallelly.
The preprocessing unit combines color images and received power of each STA to extract the features
 and sends them to the ML unit.
Additionally, the preprocessing unit computes a label, predicts the future received power effectively,
 and sends feature--label pairs to the ML unit.
The details of the feature extraction and labeling are described in section~\ref{subsec_dataprocess}. 

The ML unit holds $N$ ML models; $i$-th model predicts the future received power of the $i$-th STA.
Each ML model predicts the future received power every time the feature is received.
To train ML models, the ML unit contains $N$ training dataset queues. 
The $i$-th queue contains a certain number of feature--label pairs dataset to train $i$-th model.

In our system model, the prediction of $N$ ML models are processed in parallel,
whereas the training of each ML model is processed sequentially;
there are two reasons for such a design.
First, prediction requires more immediacy than training,
 since the prediction delay prevents real-time control of wireless links while the training is conducted in the background. 
Second, model training requires huge computation, and computation resources can be occupied by model training if conducting training multiple models parallelly.
The ML unit sequentially trains the ML model using the dataset in its training dataset queue.
If a model is still in the training process, and the other model attempts to start training,
the latter cannot start until the former has finished its training.

\subsection{Data Processing and Labeling}
\label{subsec_dataprocess}
\begin{figure}[t!]
  \centering
  \includegraphics[keepaspectratio,width = 0.8\linewidth]{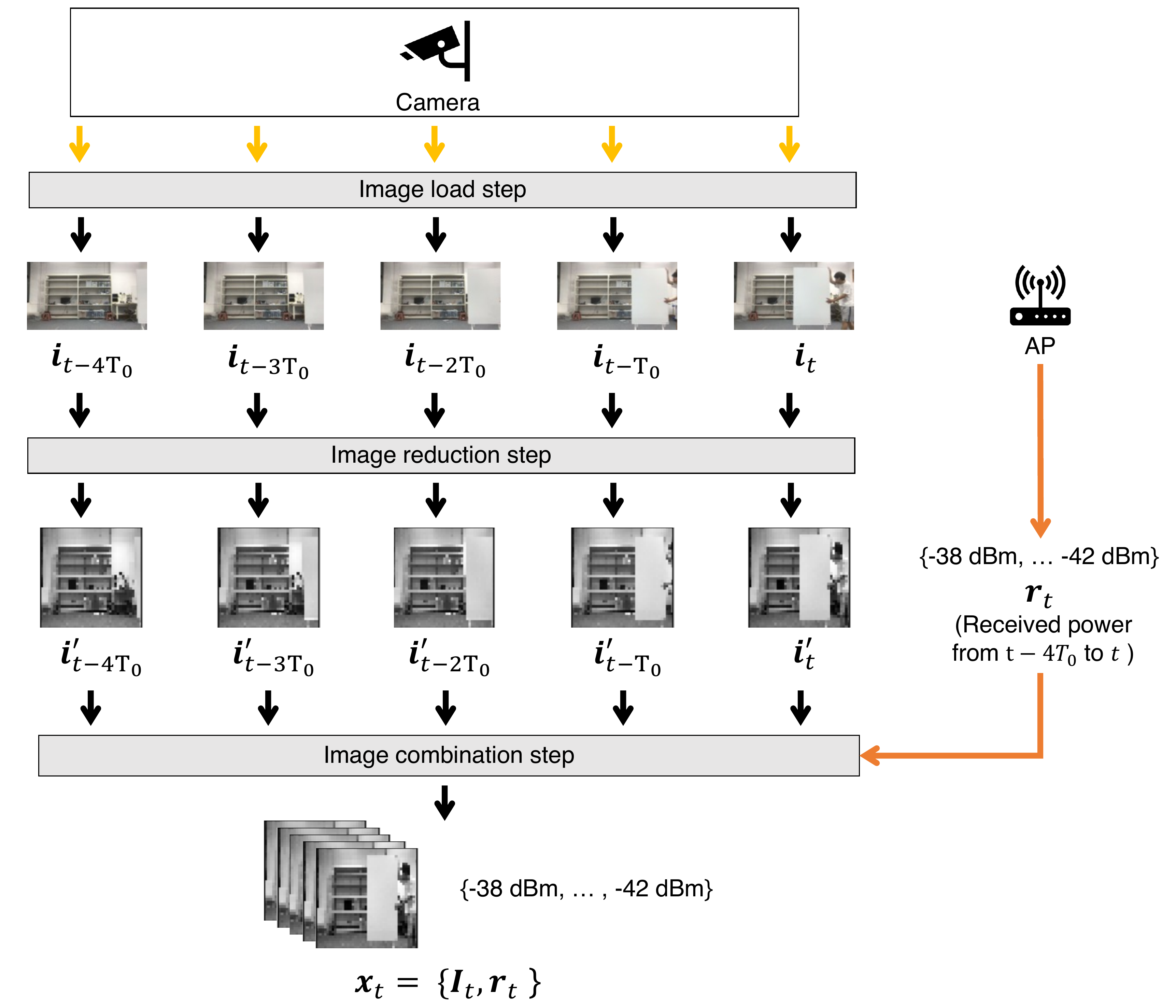}
  \caption{Feature extraction process.}
  \label{fig:feature}
\end{figure}

The feature extraction and labeling method are based on~\cite{nishio2019proactive,koda2020communication}.
In this section, we aim to (1) analyze the irradiated power of an STA,
 (2) present the feature extraction and labeling method,
  which is designed to predict the future STA received power.
In the implemented system, the feature extraction and labeling are conducted parallelly for multiple STAs.
 
We consider that the camera unit sends $F$ images per second,
 the preprocessing unit receives $F$ images and sends $F$ feature samples per second to the ML unit.
The AP sends the value of the received power of a frame transmitted by an STA to the preprocessing unit;
 let the received power at $t$ be $r_t$.
In the preprocessing unit, the images and received power values of the STA are used as input to generate a feature sample.

As shown in Fig.~\ref{fig:feature}, our feature extraction includes three steps: image loading, image reduction, and image combination.
First, let $\bm{i}_t$ be the $W \times H$ RGB image, which the camera unit captured at time $t$.
In the image loading step, the RGB image $\bm{i}_t$ is loaded from the camera unit to the preprocessing unit.
In the image reduction step, the RGB image is reduced to $w \times w$ gray image to reduce the computation cost;
 let $\bm{i}'_t$ be the gray image reduced from $\bm{i}_t$.

In the image combination step, $N_\mathrm{img}$ gray images and $N_\mathrm{r}$ received power values are combined to generate a feature sample.
First, $N_\mathrm{img}$ gray images, captured in the interval $t-(N_\mathrm{img}-1)T_0$ to $t$, are combined to form
$\bm{I}_t = \left[ \bm{i}'_{t-(N_\mathrm{img}-1)T_0} \cdots \bm{i}'_t \right]$,
where $T_0$ is time interval of images consisting a feature. 
Next, after sampling the received power captured in the interval from $t-(N_\mathrm{img}-1)T_0$ to $t$,
 with the sampling frequency of $T_0(N_\mathrm{img}-1)/N_\mathrm{r}$,
  we obtain the sequence of $N_\mathrm{r}$ received power values $\bm{r}_t = \{r_{t-(N_\mathrm{img}-1)T_0}\cdots r_{t}\}$.
Subsequently, $\bm{I}_t$ and $\bm{r}_t$ are stacked and reshaped to 1-dimensional (1D) array, which results in a feature sample $\bm{x}_t$.
The feature sample $\bm{x}_t$ is $w^2N_\mathrm{img} + N_\mathrm{r} $ dimension vector and 
 carries geometric information up to $(N_\mathrm{img}-1)T_0$ seconds ago.
The feature sample $\bm{x}_t$ is generated every time the image $\bm{i}_t$ received;
finally, $\bm{x}_t$ is sent to the ML unit.

Let us consider predicting the received power in the future $T_\mathrm{f}$ seconds.
 The feature vector $\bm{x}_t$ is associated with $r_{t+T_\mathrm{f}}$,
  resulting in the training dataset $\bm{D}_t = \{ \bm{x}_t,r_{t+T_\mathrm{f}}\}$.
The training dataset $\bm{D}_t$ is sent to the training unit.
The ML unit stores the training dataset in a training queue,
 which can contain $N_\mathrm{q}$ training samples;
  hence, the dataset generated in recent $N_\mathrm{q}/F$ seconds is used for the training.
The parameters of feature extractions and labeling are summarized in Table~\ref{tab:param_feature}.

\begin{table}[t!]
  \centering
  \caption{Parameters of feature extractions and labeling}
  \begin{tabular}{cc}
    \toprule
    Name of parameters & Value\\
    \midrule
    Camera frame rate $F$& 10 fps\\
    Row RGB image size $W \times H$& $1280\times 720$\\
    Reduced gray image size $w\times w$ & $40\times 40$\\
    Num. images of a feature sample $N_\mathrm{img}$ & 5 \\
    Time interval of each image $T_\mathrm{o}$ & 0.5 s \\
    Num. received power value of a feature sample $N_\mathrm{r}$ & 21 \\
    Dimension of a feature sample $\bm{x}_t$& 8021\\
    How far future the system predicts $T_\mathrm{f}$& 1.0 s\\
  \bottomrule
  \end{tabular}
  \label{tab:param_feature}
\end{table}

\subsection{Machine Learning Training Algorithm}

Existing studies~\cite{nishio2019proactive,koda2019handover} employed conventional long short-term memory (LSTM) networks (ConvLSTM) because ConvLSTM performs well on ML tasks with spatio-temporal features such as moving images.
However, the graphics processing unit (GPU) and enormous computation resources are required for quickly running ConvLSTM.
In this research, the intention was to implement the system with minimal computation resources available in the edge server. 
 The ML model must work quickly using only such computation resources, without GPU.

For the ML model, we used light gradient boosting machine (LightGBM)~\cite{LightGBM2017Ke}, 
which is an efficient implementation of the gradient boosting regression tree (GBRT)~\cite{friedman2001greedy}.
GBRT works by sequentially adding predictors to an ensemble, each one correcting its predecessor.
Instead of tweaking the instance weight at every iteration, GBRT tries to fit the new predictors to the residual errors made by the previous predictors.
LightGBM can run fast without GPU; it is reported that GBRT and LightGBM perform well on image-based ML tasks~\cite{reade2015hybrid}.
Therefore, we expect that the LightGBM can capture the image features and predict the received power from images.

As a first step in the ML model training process,
 each training dataset is randomly divided into two categories: updating and validation datasets, obeying the 8:2 ratio.
The model is updated using only the updating dataset for the multiple boosting rounds.
For each round, the model is evaluated on the validation dataset.
The training terminates when pre-determined boosting rounds are performed or when the validation loss is increased after several rounds consecutively, which indicates that the model has started to overfit.

\section{Experimental Setup}
\label{sec:setup}
\subsection{Implemented System}

The implemented system includes two Tx nodes as STAs, an Rx node as an AP, a camera, and an edge server.
We used a laptop equipped with a camera as the Rx, camera, and edge server.
For easy received power measurements, Wi-Fi routers were used as the Tx nodes, which transmitted beacons periodically.
The devices used for the experiment were described in Table~\ref{tab:equipment}.
The Tx nodes transmitted IEEE 802.11 beacon frames every 0.1 seconds at the same channel on the 5G band.
The Rx node measured the received power of the frame irradiated by a Tx node and sent the measured value to the server.
The camera sends the RGB image to the server every $0.1$ seconds, i.e., achieves the frame rate of 10 fps.

For the model training, we used the training queue size $T_\mathrm{q} = 50\,\mathrm{s}$,
 which means that the training queue contains the dataset generated in recent $50\,\mathrm{s}$.
The hyperparameters of the ML model training process are displayed in Table~\ref{tab:ML_HP}.

\begin{table}[t!]
  \centering
  \caption{Experimental equipment.}
  \begin{tabular}{cc}
    \toprule
    Txs& WZR-HP-AG300H\\
    \multirow{3}{*}{Laptop}  & MacBook Pro (2018) \\
    &Processor: $2.3\,\mathrm{GHz}$ quad-core Intel Core i5\\
    &Memory: $8\,\mathrm{GB}$ $2133\,\mathrm{MHz}$ $\mathrm{LPDDR3}$\\
  \bottomrule
  \end{tabular}
  \label{tab:equipment}
\end{table}

\begin{table}[t!]
  \centering
  \caption{Hyperparameters of the ML model training}
  \begin{tabular}{ccc}
    \toprule
    Hyperparameter & Value\\
    \midrule
    Num. leaf& 100\\
    Maximum depth of the tree & 8\\
    Splitting criterion & Root raised mean squared error\\
    Num. boosting round & 10\\
    Num. early stopping round & 2\\
  \bottomrule
  \end{tabular}
  \label{tab:ML_HP}
\end{table}

\subsection{Experimental Environment}

\begin{figure}[t!]
  \centering
  \begin{tabular}{c}
    \begin{minipage}[t]{0.48\hsize}
    \centering
    \includegraphics[keepaspectratio,width = 0.9\textwidth]{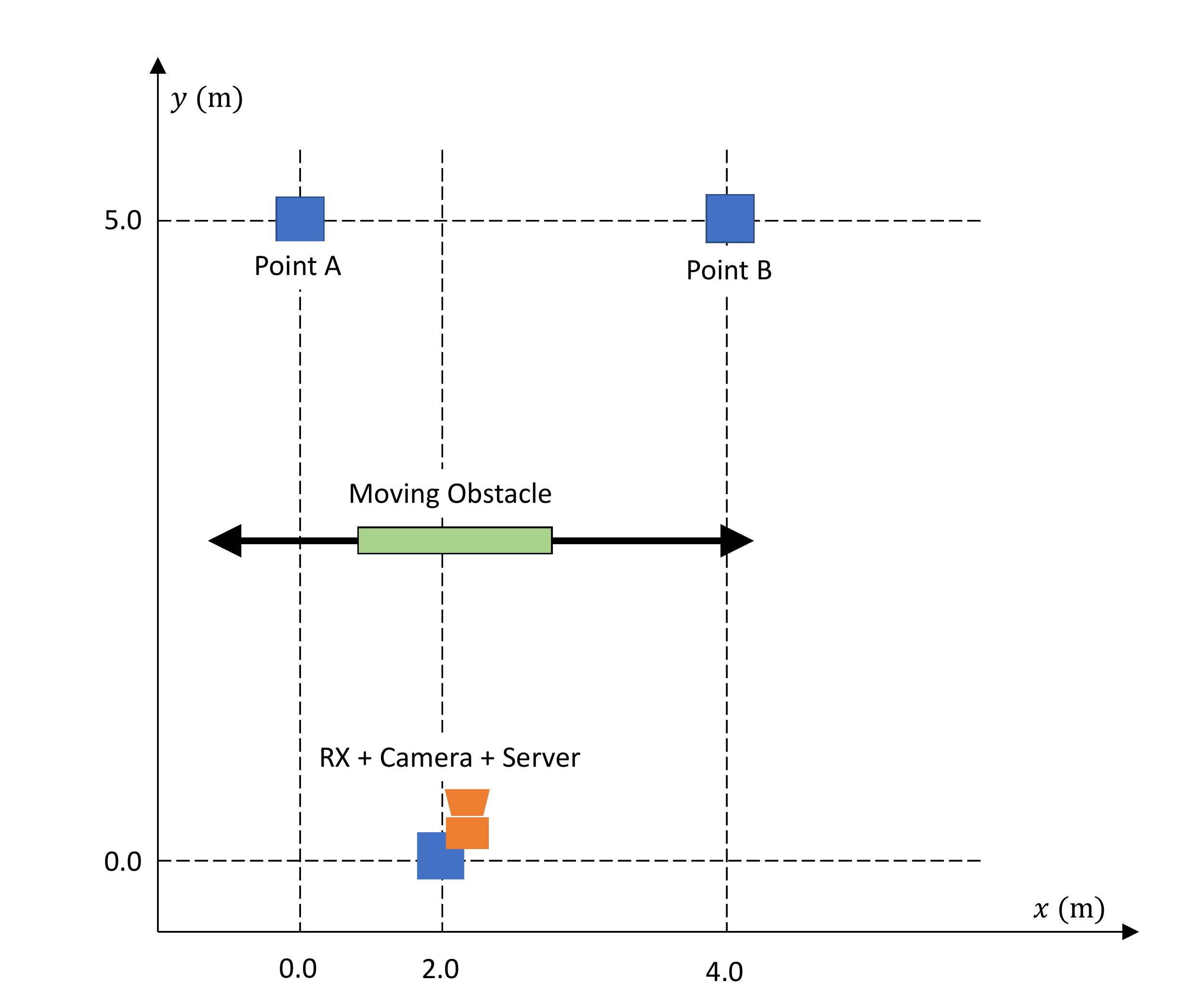}
    \subcaption{Schematics of the experimental environment for an outdoor location.}
    \end{minipage}
    \begin{minipage}[t]{0.48\hsize}
    \centering
    \includegraphics[keepaspectratio,width = 0.9\textwidth]{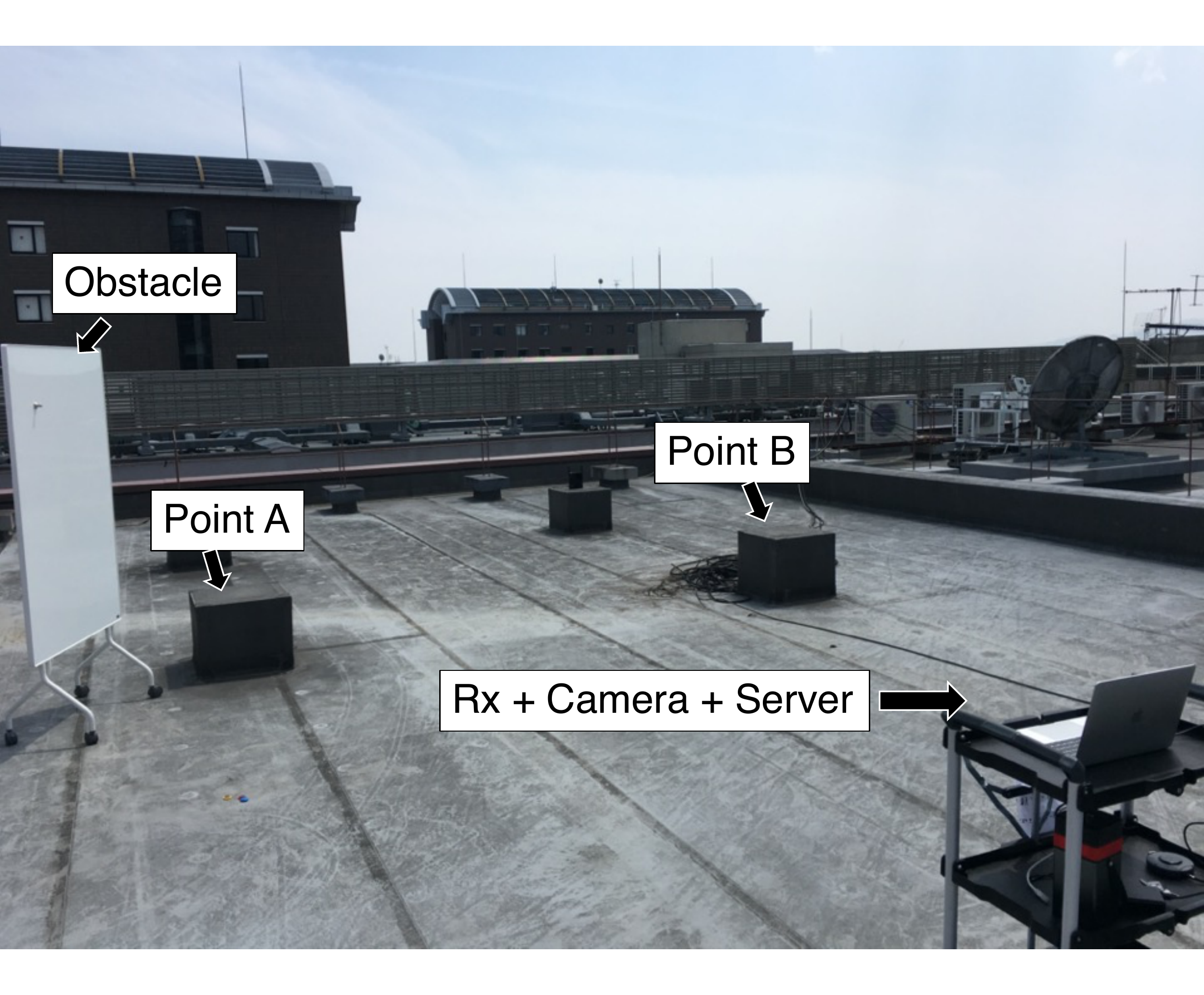}
    \subcaption{Overview of the experimental environment for an outdoor location.}
    \end{minipage}
  \end{tabular}
  \caption{Settings of an outdoor environment. The height of Point A and Point B is $0.25\,\mathrm{m}$ and that of Rx node is $0.75\,\mathrm{m}$}
  \label{fig:outdoor_env}
\end{figure}

\begin{figure}[t!]
  \centering
  \begin{tabular}{c}
    \begin{minipage}[t]{0.48\hsize}
    \centering
    \includegraphics[keepaspectratio,width = 0.9\textwidth]{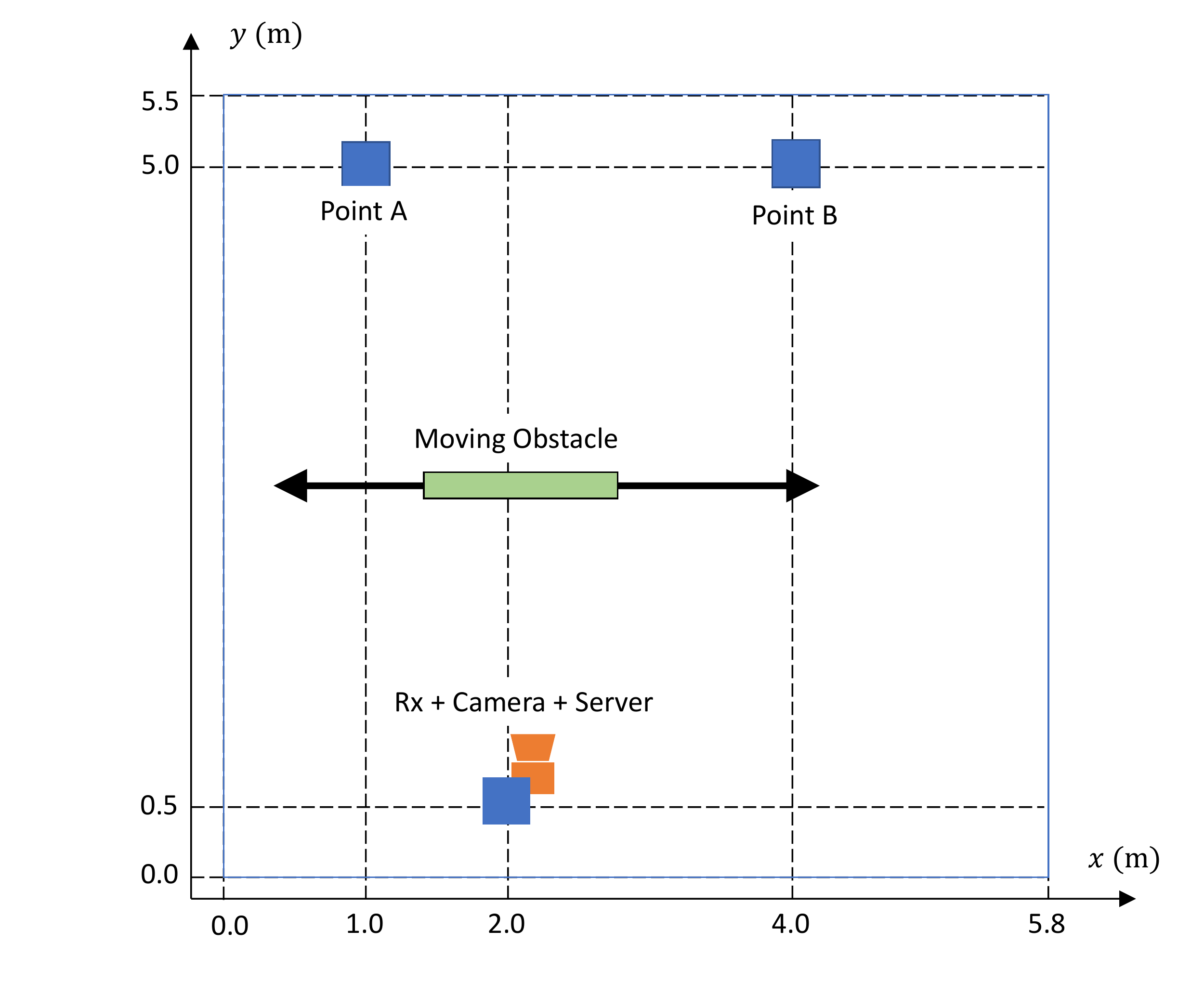}
    \subcaption{Schematics of experimental environment for an indoor location.
    Blue solid lines indicate the walls.}
    \end{minipage}
    \begin{minipage}[t]{0.48\hsize}
    \centering
    \includegraphics[keepaspectratio,width = 0.9\textwidth]{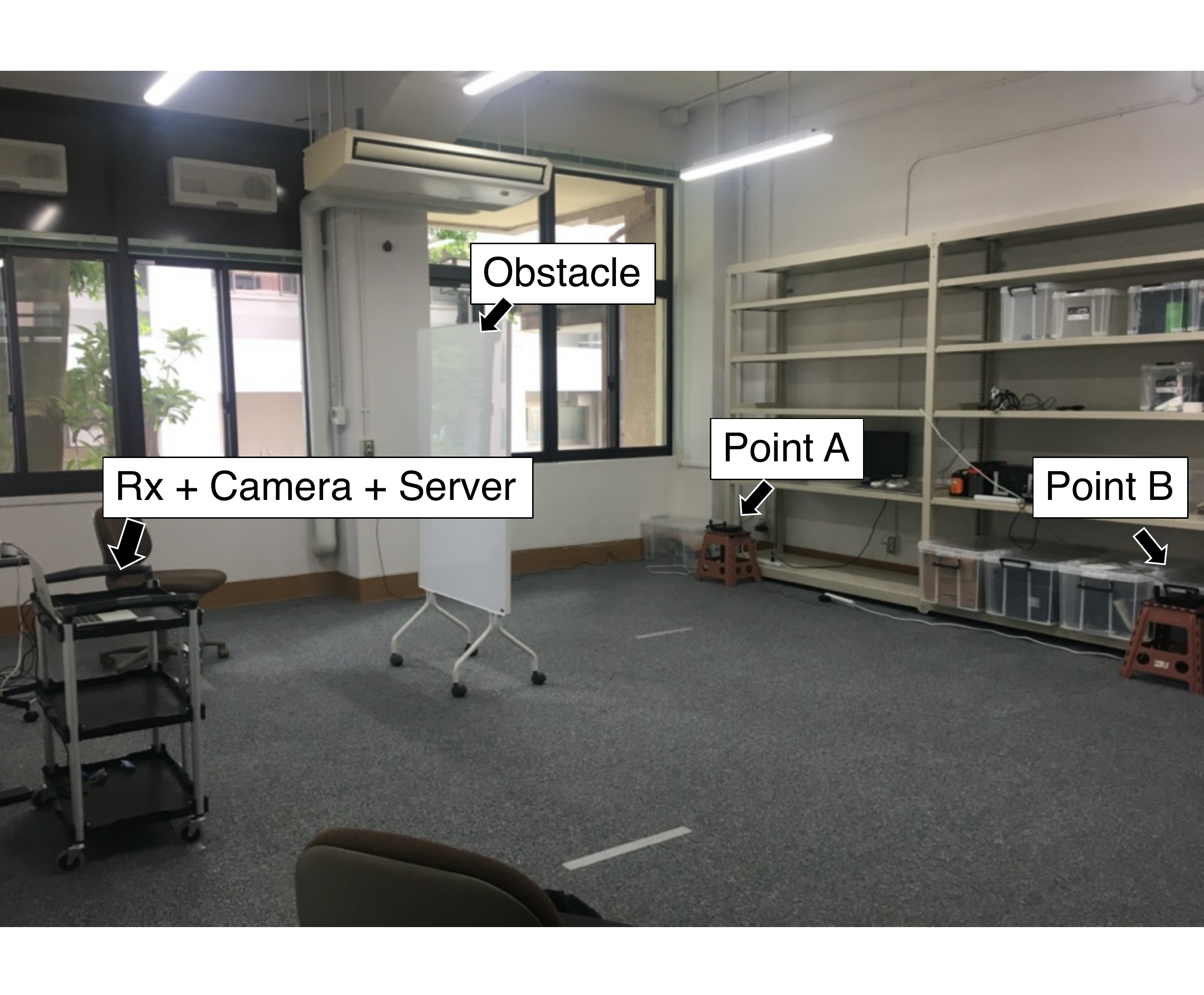}
    \subcaption{Overview of experimental environment for an indoor location.}
    \end{minipage}
  \end{tabular}
  \caption{Settings of indoor environment. The height of Point A and Point B is $0.35\,\mathrm{m}$ and that of Rx nodes is $0.75\,\mathrm{m}$}
  \label{fig:indoor_env}
\end{figure}

The experiments were conducted in two locations: indoor and outdoor.
The setups in the locations are shown in Fig~\ref{fig:outdoor_env} and~\ref{fig:indoor_env}, respectively.
According to the Japanese 5 GHz band regulation, for outdoor and indoor locations, wireless channel 100 (5.5 GHz) and 36 (5.18 GHz) were used, respectively.
The Tx nodes were placed at either Point A or Point B.
A human moved an obstacle along the $x$-axis between the Tx nodes and Rx node, and they attenuated signals.
The obstacle was a movable metal panel, with height and width of $1.7\,\mathrm{m}$ and $0.9\,\mathrm{m}$, respectively. 

\subsection{Experimental Scenario}

For both locations, two experimental scenarios were applied: stationary and mobile scenarios,
 where Tx nodes were stationary and moving, respectively.
The mobility pattern is summarized in Table~\ref{tab:scenarios}.
In the stationary scenario, positions of the Tx nodes were fixed during the experiment.
Conversely, in the mobile scenario, positions of the Tx nodes were exchanged at $300\,\mathrm{s}$,
 i.e., Tx0 was moved from Point A to Point B, and Tx1 was moved from Point B to Point A at $300\,\mathrm{s}$.
Regardless of location or scenario, the experiment took 600 seconds to complete.

\begin{table}[t!]
  \caption{Positions of Tx nodes in stationary or mobile scenarios.}
  \centering
  \begin{tabular}{cccc}
    \toprule
    \multirow{2}{*}{Scenario}&\multirow{2}{*}{Tx nodes}&\multicolumn{2}{c}{Time}\\ \cmidrule(r){3-4}
    &&$0\,\mathrm{s}-300\,\mathrm{s}$&$300\,\mathrm{s}-600\,\mathrm{s}$\\
    \midrule
    \multirow{2}{*}{Stationary}&Tx0&Point A&Point A \\
    &Tx1&Point B&Point B\\
    \multirow{2}{*}{Mobile}& Tx0&Point A&Point B \\
    &Tx1&Point B&Point A\\
  \bottomrule
  \end{tabular}
  \label{tab:scenarios}
\end{table}

\subsection{Prediction Methods with Different Features}
\label{ssec:four_methods}
This study compared three ML-based prediction methods with different features and a native prediction method.
The first ML-based prediction method uses received power and images described in Sect.~\ref{subsec_dataprocess}.
We denote this method to be received power and image-based prediction method (RP-Im), which comprises the primary method of this study.
The second ML-based prediction method uses received power, which is the received power part of the RP-Im features.
We denote this to be received power method (RP).
The third one uses the images, which is the image part of the features of RP-Im method.
We denote this to be the image method (Im).
The last is the native prediction method, which predicts that the future received power is the current one.
A more frequent blockage implied
  a higher prediction error of the native prediction method.

\section{Experimental Results}
\label{sec:results}
\subsection{Computation Time}

Table~\ref{tab:comp_time} shows the measured computation time required for the future received power prediction of the RP-Im method for each step involved in the experiment.
As described in~\ref{subsec_dataprocess}, four steps, i.e., image load, image reduction, image combination, and ML prediction, are required to make a prediction.
"Total" is the sum of computation times for these four steps,
 which denotes the latency from the time when the camera captured an image to the time when the future received power is predicted.

Average total step time is less than $0.1\,\mathrm{s}$;
this is sufficiently short to predict $1\,\mathrm{s}$ future received power.
Looking at the computation time for each step,
the image load and image reduction step are required $90\%$, 
 whereas ML prediction consumed less than $2\%$ of total step time.
Thus, to reduce total latency more, we must improve the image processing 
 by hardware acceleration or GPU.

\begin{table}[t!]
  \centering
  \caption{Average computation time of each process.}
  \begin{tabular}{ccc}
    \toprule
    Step & Average computation time $(\mathrm{ms})$ \\
    \midrule
    ML prediction& 1.3\\
    Data combination& 8.2\\
    Image reduction& 30\\
    Image load& 57\\
    \hline
    Total& 96\\
  \bottomrule
  \end{tabular}
  \label{tab:comp_time}
\end{table}

\subsection{Time Series of Prediction}
\begin{figure*}[t!]
  \centering
  \begin{minipage}[b]{\hsize}
    \centering
    \includegraphics[keepaspectratio,width=0.8\textwidth]{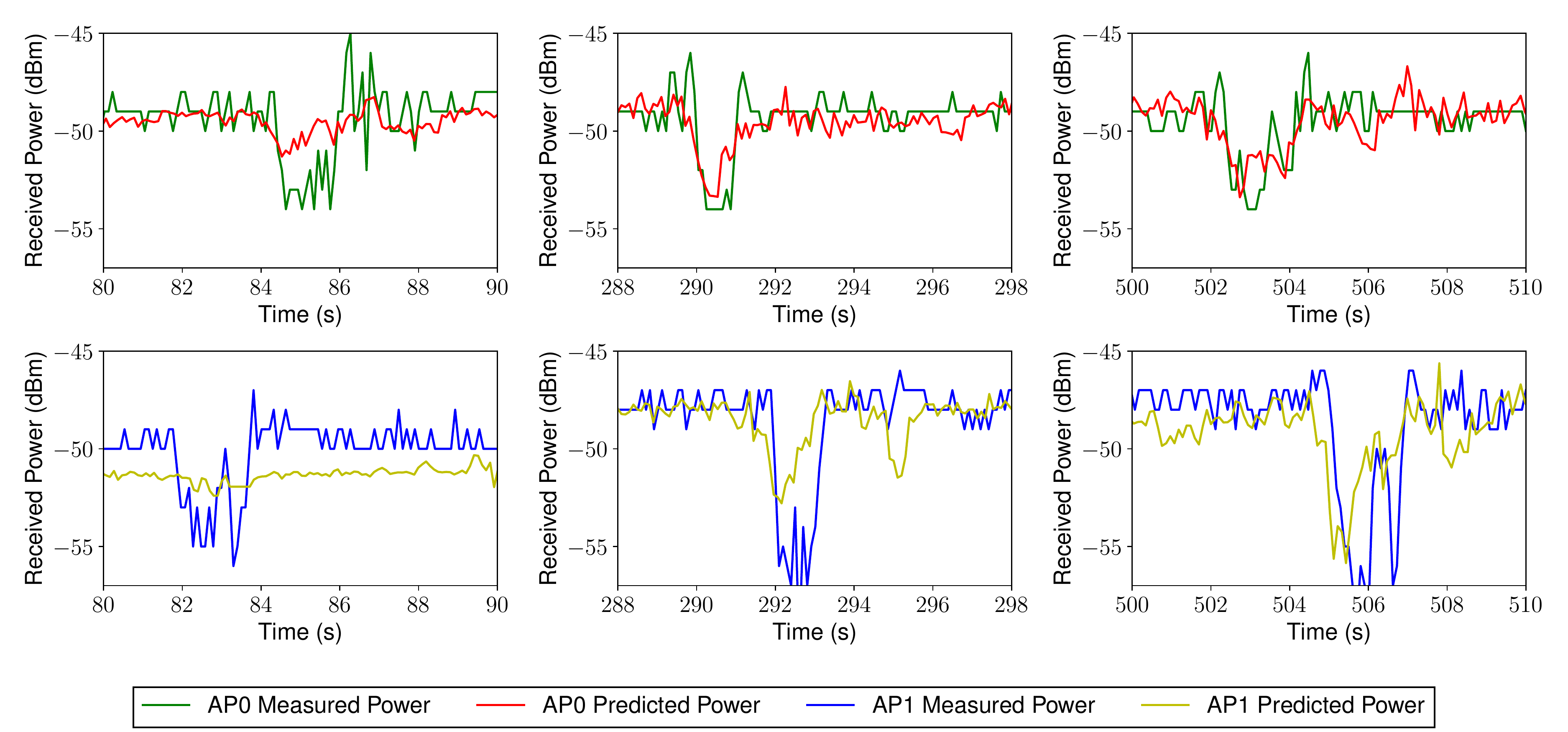}
    \caption{Predicted and measured received power of the RP-Im method on the stationary scenario at the outdoor location.
    Tx0 was at Point A and Tx1 was Point B during all the processes.
    (a) -- (c) depict results in different time slots.}
    \label{fig:stationary_pred_outdoor}
  \end{minipage}
  \begin{minipage}[b]{\hsize}
    \centering
    \includegraphics[keepaspectratio,width=0.8\textwidth]{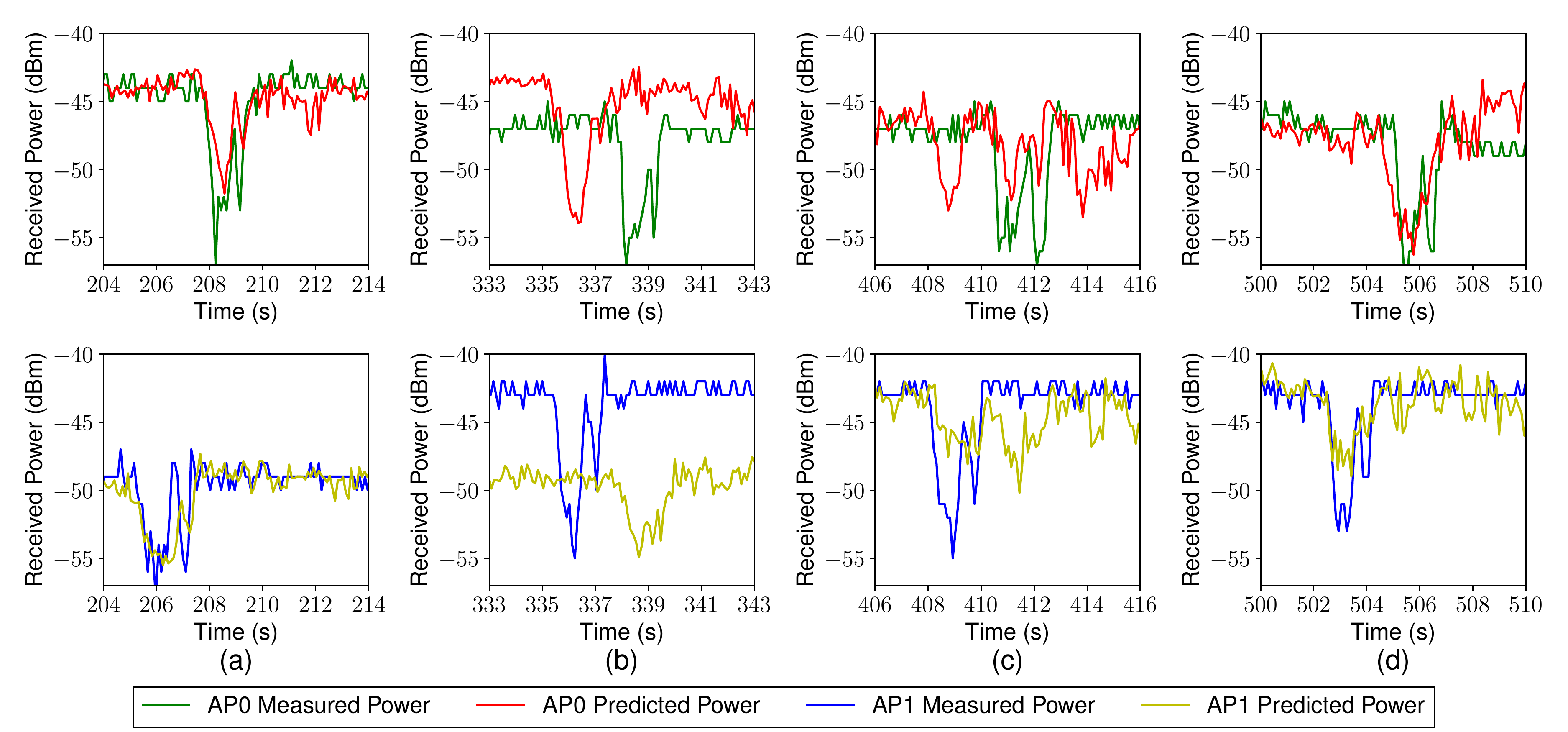}
    \caption{Predicted and measured received power of the RP-Im method on the mobile scenario at the outdoor location.
    At $300\,\mathrm{s}$, Tx0 was moved from Point A to B and Tx1 was moved from Point B to A.
    (a) -- (d) depict results in different time slots.
    }
    \label{fig:mobile_pred_outdoor}
  \end{minipage}
\end{figure*}
Figures.~\ref{fig:stationary_pred_outdoor} and~\ref{fig:mobile_pred_outdoor} depict the time series of the measured and predict the received power of the RP-Im method for two Tx nodes, 
for the stationary and mobile scenarios, respectively, for an outdoor location.
Let us focus on Fig.~\ref{fig:stationary_pred_outdoor}.
 For both Tx nodes, Tx0 and Tx1,
 the predicted received power gradually matches with the actual values as time passes.
Figure~\ref{fig:stationary_pred_outdoor}~(a) depicts the results between the $80\,\mathrm{s}$ and $90\,\mathrm{s}$, the predicted received power does not match the measured power.
This can be explained by the fact that it takes time to collect training data required for improving the model.
Figure~\ref{fig:stationary_pred_outdoor}~(c) depicts the results between the $500\,\mathrm{s}$ and $510\,\mathrm{s}$,
 the predicted received power matches the measured power much better
  because the model was trained using a larger dataset.

Figure~\ref{fig:mobile_pred_outdoor}~(a) depicts the received power between 204 s and 214 s where the model was well trained; 
 the predicted received power matches the measured power.
Figure~\ref{fig:mobile_pred_outdoor}~(b)
 depicts the results between 333 s and 343 s where the positions of the Tx nodes were exchanged;
 the predicted power substantially differs from the measured power for each.
However, the predicted power for Tx0 looks similar to the measured power for Tx1, and vice versa.
This is because the models were not trained well and not adapted to the new environment where the positions of Tx nodes were changed.
This is possible because the positions of Tx0 and Tx1 were exchanged at $300\,\mathrm{s}$, as prescribed in the mobile scenario.
As shown in Figs.~\ref{fig:mobile_pred_outdoor}~(c) and~(d), the models were gradually adapted to the Tx positions and provided more accurate received power prediction.

\subsection{Prediction Accuracy Comparison with Various Prediction Methods}

\begin{table*}[t!]
  \captionsetup{labelformat=empty,labelsep=none}
  \caption{Effect of the input feature on the prediction error (dB) in mobile scenario of Tx0.
  Gray shaded columns indicate the best performing method at each time.}
  \captionsetup{labelformat=simple,labelsep=colon}
  \begin{minipage}[t]{.45\textwidth}
    \centering
    \caption{Indoor location}
      \begin{tabular}{cccccc}
        \toprule
        Time&Native prediction&RP-Im&Im&RP\\
        \midrule
        $100\,\mathrm{s}\sim 200\,\mathrm{s}$&
        1.75&  1.04&  \cellcolor[gray]{0.8}1.03& 1.30\\ 
        $200\,\mathrm{s} \sim 300\,\mathrm{s}$&
        1.52& \cellcolor[gray]{0.8}0.92& 0.96&  1.15\\ 
        $300\,\mathrm{s} \sim 400\,\mathrm{s}$&
        \cellcolor[gray]{0.8}2.66& 2.95&  2.89&  2.71\\ 
        $400\,\mathrm{s} \sim 500\,\mathrm{s}$&
        2.10& \cellcolor[gray]{0.8}1.55& 1.58& 1.89 \\ 
        $500\,\mathrm{s} \sim 600\,\mathrm{s}$&
        2.63&  \cellcolor[gray]{0.8}1.41& 1.43&2.03 \\
        \bottomrule
      \end{tabular} 
    \label{tab:error_feature_indoor}
  \end{minipage}
  \hfill
  \begin{minipage}[t]{.45\textwidth}
    \centering
    \caption{Outdoor location}
      \begin{tabular}{cccccc}
        \toprule
        Time&Native prediction&RP-Im&Im&RP\\
        \midrule
        $100\,\mathrm{s}\sim 200\,\mathrm{s}$&
        3.28&  \cellcolor[gray]{0.8}2.03&  2.13& 2.55 \\ 
        $200\,\mathrm{s} \sim 300\,\mathrm{s}$&
        3.20& 1.84&  \cellcolor[gray]{0.8}1.82&  2.53 \\ 
        $300\,\mathrm{s} \sim 400\,\mathrm{s}$&
        \cellcolor[gray]{0.8}2.87&  3.56& 3.32&  3.31 \\ 
        $400\,\mathrm{s} \sim 500\,\mathrm{s}$&
        3.22&  2.35&  \cellcolor[gray]{0.8}2.30&  2.47\\ 
        $500\,\mathrm{s} \sim 600\,\mathrm{s}$&
        3.24&  \cellcolor[gray]{0.8}2.34& 2.38&  2.41\\ 
        \bottomrule
      \end{tabular} 
    \label{tab:error_feature_outdoor}
  \end{minipage}
\end{table*}

\begin{table*}[t!]
  \centering
  \captionsetup{labelformat=empty,labelsep=none}
  \caption{Effect of training queue size on the prediction error (dB),
  evaluated on mobile scenario of Tx0.
  Gray shaded columns indicate the best performing method at each time.
  }
  \captionsetup{labelformat=simple,labelsep=colon}
  \begin{minipage}[t]{0.45\textwidth}
    \centering
    \caption{Indoor location}
    \begin{tabular}{cccc}
      \toprule
      Time&\multicolumn{3}{c}{$T_\mathrm{q}$}\\
      &\multicolumn{1}{c}{$15\,(\mathrm{s})$}&$50\,(\mathrm{s})$&\multicolumn{1}{c}{$\infty$}\\
      \midrule
      $100\,\mathrm{s}\sim 200\,\mathrm{s}$&
       1.24& 1.04& \cellcolor[gray]{0.8}1.03\\ 
      $200\,\mathrm{s} \sim 300\,\mathrm{s}$&
       1.11& 0.92& \cellcolor[gray]{0.8}0.87 \\ 
      $300\,\mathrm{s} \sim 400\,\mathrm{s}$&
      \cellcolor[gray]{0.8}2.57& 2.95& 2.81\\ 
      $400\,\mathrm{s} \sim 500\,\mathrm{s}$&
       \cellcolor[gray]{0.8}1.52& 1.55& 1.68\\ 
      $500\,\mathrm{s} \sim 600\,\mathrm{s}$&
       1.53& \cellcolor[gray]{0.8}1.41& 1.45\\
      \bottomrule
    \end{tabular} 
    \label{tab:error_buffer_indoor}
  \end{minipage}
  \hfill
  \centering
  \begin{minipage}[t]{0.45\textwidth}
    \centering
    \caption{Outdoor location}
    \begin{tabular}{cccc}
      \toprule
      Time&\multicolumn{3}{c}{$T_\mathrm{q}$}\\
      &\multicolumn{1}{c}{$15\,(\mathrm{s})$}&$50\,(\mathrm{s})$&\multicolumn{1}{c}{$\infty$}\\
      \midrule
      $100\,\mathrm{s}\sim 200\,\mathrm{s}$&
       2.16& 2.03& \cellcolor[gray]{0.8}1.79\\
      $200\,\mathrm{s} \sim 300\,\mathrm{s}$&
      1.84& 1.84& \cellcolor[gray]{0.8}1.79 \\ 
      $300\,\mathrm{s} \sim 400\,\mathrm{s}$&
      \cellcolor[gray]{0.8}3.56& \cellcolor[gray]{0.8}3.56& \cellcolor[gray]{0.8}3.56\\ 
      $400\,\mathrm{s} \sim 500\,\mathrm{s}$&
       3.18& \cellcolor[gray]{0.8}2.35& 2.49\\ 
      $500\,\mathrm{s} \sim 600\,\mathrm{s}$&
       2.83& 2.34& \cellcolor[gray]{0.8}2.28\\ 
      \bottomrule
    \end{tabular} 
    \label{tab:error_buffer_outdoor}
  \end{minipage}
  \label{tab:error_buffer}
\end{table*}

Tables~\ref{tab:error_feature_indoor} and~\ref{tab:error_feature_outdoor}
 show the comparison of four prediction methods described in~\ref{ssec:four_methods} in terms of prediction error in the mobile scenario,
 where the Tx nodes were exchanged at $300\,\mathrm{s}$.
On comparing the naive prediction and three ML-based methods with different features, 
 before $300\,\mathrm{s}$, we find that the
 three ML-based methods outperform the native prediction method in terms of prediction error.
This result shows that, without the change of the environment, the ML-based predictions are more accurate than that of native method.
From $300\,\mathrm{s}$ to $400\,\mathrm{s}$ (immediately after the environment change), the
three ML-based methods perform inferior or equivalent to the native prediction method.
This means that from the time at which the environment changes until the time at which the model adapts to this change, 
the prediction error of the ML-based methods is higher than that of the native prediction method.
However, between $400\,\mathrm{s}$ and $600\,\mathrm{s}$, the three ML-based methods outperformed the native prediction method.
After the model adapts to the environment change, the ML-based methods outperform native method.
The time taken to adapt to the new environment is about $100\,\mathrm{s}$
because the ML-based methods outperform the native method at a time between $400\,\mathrm{s}$ and $500\,\mathrm{s}$.
This time is unacceptably long for a mobile communication system.
However, this is sufficiently short for different scenarios, such as mm-wave wireless lead-in line to indoor.

The comparison of the ML-based methods with different features shows that the
 prediction errors of the RP-Im and Im methods are similar, and both are lower than that of the RP method.
As shown by existing studies~\cite{nishio2019proactive,koda2019handover,koda2020communication},
an image was more informative than other features for the future received power prediction.

\subsection{Prediction Accuracy for Various Training Queue Size}

Tables~\ref{tab:error_buffer_indoor} and~\ref{tab:error_buffer_outdoor} show the effect of the training queue size in terms of prediction error on the mobile scenario using the RP-Im method.
Considering the training queue size, the following fact is worth noting: for a larger training queue size, the data points that populate the dataset used for the training are older.
For example, taking the training queue size $N_\mathrm{q}$ and camera frame rate $F\,(\mathrm{fps})$,
the dataset generated from $N_\mathrm{q}/F$ seconds ago to the present is used for training.
Let us denote $N_\mathrm{q}/F$ as $T_\mathrm{q}$.

Until $300\,\mathrm{s}$, when the positions of Tx nodes were changed,
 the larger $T_\mathrm{q}$ is associated with the lower prediction error because a larger training dataset is used for a larger $T_\mathrm{q}$.
Between $300\,\mathrm{s}$ and $400\,\mathrm{s}$,
 the prediction error of $T_\mathrm{q}=\infty$ is inferior or equivalent to that of $T_\mathrm{q}=\,50\mathrm{s}$ or $T_\mathrm{q}=15\,\mathrm{s}$.
Considering the training at time $t$ seconds, where $t > 300$, i.e., after the positions of Tx nodes were changed, 
 when $T_\mathrm{q}$ is larger than $t - 300$, the training dataset used include the old data that was generated before $300\,\mathrm{s}$.
Using this old dataset prevents the ML model from adapting to a new setting.
This negative effect of old data is maybe more severe for a large $T_\mathrm{t}$.

\section{Conclusion}
\label{sec:conclusion}

This study proposed a real-time online trainable wireless link quality prediction system and implemented the system with commercially available laptops.
The indoor and outdoor experiments demonstrated that the prediction model was updated in real-time, adapted to the environment change, and accurately predicted the time-varying Wi-Fi received power.
Out future works include developing online training and prediction method to learn and predict the received power for moving STAs.

\section*{Acknowledgment}
  This work was supported in part by JSPS KAKENHI Grant Numbers JP17H03266, JP18K13757 and KDDI Foundation.

\bibliographystyle{IEEEtran}
\bibliography{itahara_GCWS_cameraready}

\begin{thebibliography}{10}
\providecommand{\url}[1]{#1}
\csname url@samestyle\endcsname
\providecommand{\newblock}{\relax}
\providecommand{\bibinfo}[2]{#2}
\providecommand{\BIBentrySTDinterwordspacing}{\spaceskip=0pt\relax}
\providecommand{\BIBentryALTinterwordstretchfactor}{4}
\providecommand{\BIBentryALTinterwordspacing}{\spaceskip=\fontdimen2\font plus
\BIBentryALTinterwordstretchfactor\fontdimen3\font minus
  \fontdimen4\font\relax}
\providecommand{\BIBforeignlanguage}[2]{{%
\expandafter\ifx\csname l@#1\endcsname\relax
\typeout{** WARNING: IEEEtran.bst: No hyphenation pattern has been}%
\typeout{** loaded for the language `#1'. Using the pattern for}%
\typeout{** the default language instead.}%
\else
\language=\csname l@#1\endcsname
\fi
#2}}
\providecommand{\BIBdecl}{\relax}
\BIBdecl

\bibitem{Rappaport2019Wireless}
T.~S. {Rappaport} \emph{et~al.}, ``Wireless communications and applications
  above 100 {GHz}: Opportunities and challenges for 6{G} and beyond,''
  \emph{{IEEE} Access}, vol.~7, pp. 78\,729--78\,757, Jun. 2019.

\bibitem{collonge2004influence}
S.~Collonge \emph{et~al.}, ``Influence of the human activity on wide-band
  characteristics of the 60 {GHz} indoor radio channel,'' \emph{{IEEE} Trans.
  Wireless Commun.}, vol.~3, no.~6, pp. 2396--2406, Nov. 2004.

\bibitem{nishio2019proactive}
T.~Nishio \emph{et~al.}, ``Proactive received power prediction using machine
  learning and depth images for mmwave networks,'' \emph{{IEEE} J. Sel. Areas
  Commun}, vol.~37, no.~11, pp. 2413--2427, Nov. 2019.

\bibitem{ayvacsik2019veni}
S.~Ayva{\c{s}}{\i}k \emph{et~al.}, ``Veni vidi dixi: reliable wireless
  communication with depth images,'' in \emph{Proc. CoNEXT}, Orlando, FL, USA,
  Dec. 2019, pp. 172--185.

\bibitem{koda2019handover}
Y.~Koda \emph{et~al.}, ``Handover management for mmwave networks with proactive
  performance prediction using camera images and deep reinforcement learning,''
  \emph{{IEEE} Trans. Cogn. Commun. Netw.}, vol.~6, no.~2, pp. 802--816, Dec.
  2019.

\bibitem{iimori2020stochastic}
H.~Iimori \emph{et~al.}, ``Stochastic learning robust beamforming for
  millimeter-wave systems with path blockage,'' \emph{{IEEE} Wirel. Commun.},
  May 2020.

\bibitem{koda2020communication}
Y.~Koda \emph{et~al.}, ``Communication-efficient multimodal split learning for
  mmwave received power prediction,'' \emph{{IEEE} Commun. Lett}, vol.~24,
  no.~6, pp. 1284--1288, May 2020.

\bibitem{LightGBM2017Ke}
G.~Ke \emph{et~al.}, ``Lightgbm: A highly efficient gradient boosting decision
  tree,'' in \emph{Proc. NIPS}, Long Beach, CA, USA, Dec. 2017, pp. 3146--3154.

\bibitem{friedman2001greedy}
J.~H. Friedman, ``Greedy function approximation: a gradient boosting machine,''
  \emph{Ann. Statist.}, vol.~29, no.~5, pp. 1189--1232, Oct. 2001.

\bibitem{reade2015hybrid}
S.~Reade and S.~Viriri, ``Hybrid age estimation using facial images,'' in
  \emph{Proc. ICIAR}, Niagara Falls, Canada, Jul. 2015, pp. 239--246.

\end{thebibliography}
\end{document}